\documentclass[conference,letterpaper,twoside,final,10pt]{IEEEtran}

\usepackage{graphicx}				
\usepackage[tight]{subfigure}		
\usepackage{amssymb}				
\usepackage{amsmath}				
\usepackage{url}
\usepackage{cite}					
\usepackage{xcolor}					
\usepackage[mode=text]{siunitx} 	

\usepackage[inline]{enumitem}		
\usepackage{acronym} 				
\usepackage{microtype}				
\usepackage{tikz}					
\usepackage{pgfplots}				
\usepackage{pgfplotstable}			
\usepackage{tikzscale}				
\usepackage{helvet}					
\usepackage[eulergreek]{sansmath}	
\usepackage{algorithm}				
\usepackage[noend]{algpseudocode}	
\usepackage{tabularx}				
\usepackage{booktabs}				

\usepackage[capitalize]{cleveref}	

\usetikzlibrary{%
	 arrows%
	,automata%
	,calc%
	,fit%
	,patterns%
	,positioning%
	,shadows%
	,shapes.misc%
}

\tikzset{cross/.style={cross out, draw=black, minimum size=2*(#1-\pgflinewidth), inner sep=0pt, outer sep=0pt},
cross/.default={3pt}}

\makeatletter
\newcommand*{\org@overidelabel}{}
\let\org@overridelabel\@verridelabel
\@ifpackagelater{acronym}{2015/03/21}{
  \renewcommand*{\@verridelabel}[1]{%
    \@bsphack
    \protected@write\@auxout{}{\string\AC@undonewlabel{#1@cref}}%
    \org@overridelabel{#1}%
    \@esphack
  }%
}{
  \renewcommand*{\@verridelabel}[1]{%
    \@bsphack
    \protected@write\@auxout{}{\string\undonewlabel{#1@cref}}%
    \org@overridelabel{#1}%
    \@esphack
  }%
}
\makeatother

\floatname{algorithm}{Algorithm}

\newcommand*\Let[2]{\State #1 $\gets$ #2}
\newcommand{\myalgnumfont}{\fontsize{5pt}{10pt}\selectfont\sffamily\color{black!90!white}}
\algrenewcommand{\alglinenumber}[1]{\myalgnumfont #1:}

\Crefname{figure}{Fig.}{Figs.}
\Crefname{paragraph}{Section}{Sections}

\DeclareMathOperator\erfc{erfc}

\IEEEoverridecommandlockouts

\newtheorem{lemma}{Lemma}
\newtheorem{proposition}{Proposition}

\newtheorem{corollary}{Corollary}

\acrodef{ack}[{\texttt{ACK}}]{Acknowledgement}
\acrodef{adc}[ADC]{Analog to Digital Converter}
\acrodef{api}[API]{Application Programming Interface}
\acrodef{asm}[ASM]{Assembly}
\acrodef{ble}[BLE]{Bluetooth Low Energy}
\acrodef{blisp}[BLISP]{platform combined of \acs{ble} and \acs{wisp}}
\acrodef{cots}[COTS]{Commercial off-the-Shelf}
\acrodef{crfid}[CRFID]{Computational RFID}
\acrodef{dut}[DUT]{Device Under Test}
\acrodef{epcc1g2}[EPC~C1G2]{EPCglobal Class\,1 Generation\,2}
\acrodef{epc}[{\texttt{EPC}}]{Electronic Product Code}
\acrodef{fram}[FRAM]{Ferroelectric Random Access Memory}
\acrodef{gpio}[GPIO]{General Purpose Input/Output}
\acrodef{gsm}[GSM]{Global System for Mobile communication}
\acrodef{ide}[IDE]{Integrated Developement Environment}
\acrodef{iot}[IoT]{Internet of Things}
\acrodef{llrp}[LLRP]{Low-Level Reader Protocol}
\acrodef{lsb}[LSB]{Least Significant Bit}
\acrodef{nfc}[NFC]{Near Field Communication}
\acrodef{ocr}[OCR]{Optical Character Recognition}
\acrodef{pcb}[PCB]{Printed Circuit Board}
\acrodef{rfid}[RFID]{Radio Frequency Identification}
\acrodef{rf}[RF]{Radio Frequency}
\acrodef{rn16}[{\texttt{RN16}}]{16bit Random Number}
\acrodef{rssi}[RSSI]{Received Signal Strength Indication}
\acrodef{sdk}[SDK]{Software Development Kit}
\acrodef{soc}[SoC]{System on Chip}
\acrodef{spi}[SPI]{Serial Peripheral Interface}
\acrodef{twi}[TWI]{Two Wire Interface}
\acrodef{uart}[UART]{Universal Asynchronous Receiver/Transmitter}
\acrodef{usb}[USB]{Universal Serial Bus}
\acrodef{vlc}[VLC]{Visual Light Communication}
\acrodef{wisp}[WISP]{Wireless Identification and Sensing Platform}

\newcommand{\acresetwithnoexpand}{%
	\acresetall
	\acused{gsm}
	\acused{rfid}
	\acused{rf}
	\acused{rn16}
}

\begin{document}
\title{\acs{blisp}: Enhancing Backscatter Radio with Active Radio for \aclp{crfid}}
\author{
	\IEEEauthorblockN{Ivar in\,'t Veen\IEEEauthorrefmark{1}, Qingzhi Liu\IEEEauthorrefmark{1}, Przemys{\l}aw Pawe{\l}czak\IEEEauthorrefmark{1}, Aaron Parks\IEEEauthorrefmark{2}, and Joshua R. Smith\IEEEauthorrefmark{2}}
	\IEEEauthorblockA{
		\IEEEauthorrefmark{1}
		Delft University of Technology, Mekelweg\,4, 2628\,CD Delft, The Netherlands\\
		Email: \{qinzhi.q.liu, p.pawelczak\}@tudelft.nl\\
	}
	\IEEEauthorblockA{
		\IEEEauthorrefmark{2}
		University of Washington, Seattle, WA 98195-2350, USA\\
		Email: anparks@uw.edu, jrs@cs.uw.edu
	}
	\thanks{Supported by the Dutch Technology Foundation STW under contract 12491 and in part by a Google Faculty Research Award, a Google PhD fellowship, the Intel Science and Technology Center for Pervasive Computing, and NSF award CNS-1305072.}
	\thanks{Another version of this work is available in~\cite{veen2015:thesis}.}
	\thanks{\copyright~2016 IEEE. Personal use of this material is permitted. Permission from IEEE must be obtained for all other uses, in any current or future media, including reprinting/republishing this material for advertising or promotional purposes, creating new collective works, for resale or redistribution to servers or lists, or reuse of any copyrighted component of this work in other works.}
}

\maketitle

\acresetwithnoexpand
\begin{abstract}
We demonstrate the world's first hybrid radio platform which combines the strengths of active radio (long range and robustness to interference) and \aclp{crfid} (low power consumption).
We evaluate the \acf{wisp}, an \acs{epcc1g2} standard-based, \acl{crfid} backscatter radio, against \acf{ble} and show (theoretically and experimentally) that \acs{wisp} in high channel attenuation conditions is less energy efficient per received byte than \acs{ble}.
Exploiting this observation we design a simple switching mechanisms that backs off to \acs{ble} when radio conditions for \acs{wisp} are unfavorable.
By a set of laboratory experiments, we show that our proposed hybrid active/backscatter radio obtains higher goodput than \acs{wisp} and lower energy consumption than \acs{ble} as stand-alone platforms, especially when \acs{wisp} is in range of an \ac{rfid} interrogator for the majority of the time.
Simultaneously, our proposed platform is as energy efficient as \acs{ble} when user is mostly out of \ac{rfid} interrogator range.
\end{abstract}
\acresetwithnoexpand

\setlength{\textfloatsep}{0pt}

\section{Introduction}
\label{sec:introduction}

Most low power wireless sensor nodes use active radio transmission techniques, such as Bluetooth Low Energy~\cite{gomez2012sensors}, to transport data.
While active radios are becoming better with each year (in terms of throughput and range), the power consumption expenditure of radio communication can still be much larger than the power expended for computation~\cite{fonseca2008quanto}.
This indicates that there is still a lot to be done to make wireless sensor nodes more power efficient, despite many years of research in low power electronics.
One approach to reducing energy consumption of the wireless front end is by not actively transmitting, but instead modulating the reflection of power emitted by an external transmitter---as with \ac{rfid}-based \acp{crfid}~\cite{sample2008im}.

\subsection{Problem Statement and Research Question}
\label{sec:introduction/problem}

Unfortunately the transmission technique used by \acp{crfid}, i.e. backscatter, has non-ideal characteristics compared to active radio.
While power efficient, backscatter is susceptible to distortion by the environment~\cite[Fig. 4]{zhang2012mobisys}.
Additionally, the path loss for backscatter signals is very different than for active transmissions.
Active transmissions have a signal-to-noise ratio which approximately decays with the square of distance.
For backscatter radio, this decay approximates the fourth power of distance~\cite[Sec. 2.2]{zhang2012mobisys}.
Hence, the energy wasted due to lost data increases.
At the same time, active radios, although more resistant to interference, consume more energy than backscatter radios.
The difference in power consumption is mostly due to the need to actively emit \ac{rf} power instead of reflecting preexisting signals.
This robustness/energy efficiency trade-off of active and backscatter radio calls for connecting these platforms.
Practically, many real-life situations call for an extension of backscatter by active radio.

\textbf{Example:} It is shown in~\cite{guo2009cows} that cows have preferred regions (hotspots) within the paddock in which they spend the majority of their time. In~\cite[Fig. 3]{guo2009cows} the number of hotspots (covering less than 20\,\si{\square\meter}) is limited to six, and is spread over a large area ($\approx$230\,\si{\square\meter}). To monitor cattle movement (C)\acs{rfid} would cover the hotspot area, while active radio would cover transitional movement.

The research question is then:
\textit{What energy consumption and transmission reliability improvements can one get by exploiting the combined benefits of active and backscatter radio?}

\subsection{Contributions of This Paper}
\label{sec:introduction/contributions}

To answer this question we design a new heterogeneous radio sensor node combining both active and passive radio in one device.
We call this platform \acsu{blisp}---a composition of \textbf{B}luetooth \textbf{L}ow Energy (state-of-the-art \ac{cots} active radio platform for consumer applications~\cite{gomez2012sensors}) and W\textbf{ISP}~\cite{wiki2015wisp5} (state-of-the-art \ac{crfid}).
This proposed platform consists both of low-cost experimental hardware combining the two radios in one system, and a radio selection technique (implemented in software) to choose the appropriate radio for the appropriate situation while trying to optimize both reliability and energy efficiency.
To show the benefit of \ac{blisp}, the complete system is evaluated in replicable static and mobile scenarios using a \ac{cots} \ac{rfid} reader and a modified smartphone-attached \ac{rfid} reader.

The contributions presented in this paper are:

\textbf{Contribution 1:} we provide a set of simple theories, supported by experiments, showing the benefit of connecting active and backscatter radio platforms;

\textbf{Contribution 2:} we show the benefit of using \ac{blisp} as an extension to \ac{crfid} applications by demonstrating that it is possible to transmit more data compared to an out-of-range \ac{crfid} while only increasing energy consumption per byte by $\approx$15\,\% compared to \ac{ble}.

\textbf{Contribution 3:} we show the benefit of using \ac{blisp} as an extension to \ac{ble} applications by demonstrating the possibility of transmitting the same amount of data compared to \ac{ble} while decreasing energy consumption per byte by more than 50\,\%.

The rest of this paper is organized as follows.
\Cref{sec:related_work} reviews related work.
Research motivation is provided in~\Cref{sec:motivation}, followed by~\Cref{sec:radio-switching} discussing a simple feedback-less radio switching method for \ac{blisp}.
\Cref{sec:blisp_design} presents experimental platform used to verify the quality of the proposed switching mechanism of which the results are discussed in~\Cref{sec:experiments}.
A discussion on limitations and future work is given in~\Cref{sec:future}, and the paper concludes with~\Cref{sec:conclusions}.

\section{Related Work}
\label{sec:related_work}

We start by reviewing literature pertaining to active and backscatter radios and connection thereof into an hybrid device.

\subsection{\acl{crfid}}
\label{sec:related_work/backscatter}

The use of \acs{crfid} for wireless sensor applications has been advocated by many papers including~\cite{yeager2008rfid,philipose2005pervasive}.
The only stable \ac{crfid}~\cite{sample2008im} implementation currently available is {~~}\ac{wisp}.
The communication protocol used by \ac{wisp} is the industrial standard \ac{epcc1g2} \ac{rfid} protocol. Although completely battery-autonomous, \ac{crfid} has intrinsic limitations:
Limited channel robustness, as evaluated by~\cite{zhang2012mobisys}; and limited \ac{rf} power transfer efficiency results in an intermittent power supply.
A solution to the continuous power supply problem proposed by~\cite{dong2015rfid} exercises a hybrid power solution based on \ac{rf} power harvesting and an energy storage device.
While this significantly improves \ac{crfid} energy supply stability, it does not solve the robustness problem.

\subsection{\acl{ble}}
\label{sec:related_work/active}

Active (low power) radio systems are less susceptible to interference compared to backscatter communication.
However, they bring the disadvantage of higher energy consumption.
There are multitudes of low power active radio platforms, and reviewing all options is not in the scope of this work.
However, there is one believed to be broadly adopted, with more than 30 billion devices expected to reach the consumer market by 2020~\cite{ble2015standard}: \ac{ble}---the newest version of the Bluetooth protocol optimized for low energy applications\footnote{For example, recent standards like SigFox~\cite{sigfox2015web}, LoRa~\cite{lora2015web} or IEEE 802.15.4k~\cite{ieee2013802154k} could be used, and are expected to have even lower energy consumption than \ac{ble}.
We will not use them in this work as they are not (yet) easily accessible for experimental evaluation, nor broadly adopted.}.
Works by~\cite{gomez2012sensors,kamath2010an092} experimentally evaluate the performance of \ac{ble}, while~\cite{siekkinen2012wcncw} shows the energy consumption of \ac{ble} compared to other popular active radio technologies.
No studies comparing the energy consumption of \ac{ble} with a backscatter-based \ac{crfid} have yet been published to the best of our knowledge.

\subsection{Multi-Radio Systems}
\label{sec:related_work/multi}

A combination of backscatter radio and active radio seems to be the logical step to solve the imperfections of both systems.
Again, to the best of our knowledge, no such hybrid implementation exists. 
One obvious way of using \ac{ble} to extend the \ac{rfid} range is to use multiple \ac{rfid} readers which are coupled using \ac{ble}, as proposed for different radio types (with node-to-node communication) by~\cite{islam2014access}.
This approach, unfortunately, cannot be used for \ac{blisp} because state of the art \ac{crfid}s cannot communicate with other \ac{crfid}s without the interrogator.
The only hybrid active/backscatter platform we are aware of is~\cite{kampianakis2014real}, which uses \ac{ble} to reprogram a backscatter testbed, and does not use the active radio to improve reliability.

Authors of~\cite{brideglall2007symbol} propose a method of using \ac{ble} as a physical transport layer for an \acs{rfid} protocol.
A backscatter-\acs{ble} method is proposed in~\cite{ensworth2015rfid}, which allows a backscatter device to synthesize \acs{ble} packets but which has similar channel constraints as conventional backscatter.
In the non-backscatter context, an approach to combine multiple heterogeneous radios by~\cite{gummeson2010jsac} uses acknowledgement delay and machine learning mechanisms to optimize system performance.
All above-mentioned multi-radio platforms rely on acknowledgements from the receiving party and/or active radio transmissions.

\subsection{\ac{rf} Power Harvesting}
\label{sec:related_work/harvester}

Considering literature related to energy storage in \ac{crfid},
we need to mention \cite{dong2015rfid} again proposing to store energy in battery/capacitor for future use and~\cite{talla2015arxiv} where energy storage from rectifying Wi-Fi signals has been proposed.

\section{Motivation for Combining Active and Backscatter Radio}
\label{sec:motivation}

To understand why backscatter is not always the most efficient radio technique, we introduce a simple analytical basis to bring insight into the design of \ac{blisp}. The theoretical model is followed by experimental results verifying the theory.

\subsection{Difference in \acs{wisp} and \acs{ble} Radio Efficiency}
\label{sec:motivation/efficiency}

We start with the analytical model.

\subsubsection{Analysis}
\label{sec:motivation/evaluation}

Assume a hybrid radio platform composed of $i=\{1,\ldots,n\}$ independent radio technologies (such as backscatter and active radio). We characterize the energy per successful transferred byte for radio $i$ as $E_{\text{byte},i}(d) = E_{\text{tx},i} / B_{\text{rx},i}(d)$, where $E_{\text{tx},i}$ is the total amount of energy spent in transmitting data and $B_{\text{rx},i}(d)$ is the number of received bytes for distance $d \in [0, d_{\max})$. Generalizing~\cite[Sec. III-A]{lettieri1998info}
\begin{equation}
	B_{\text{rx},i}(d) \triangleq \frac{L}{L+H} \left[ 1 - \erfc\left( f_{i}(d) \right) \right] ^ {L+H},
	\label{eq:rx_bytes}
\end{equation}
where $L$ and $H$ are the payload size in bits and the amount of overhead in bits, respectively, and $\erfc(.)$ is the complementary error function.
We define the signal quality decay function $f_{i}(d) = \left( d / a_{i} \right) ^ {-r_{i}}$, $a_{i}$ as radio-intrinsic correction value and $r_{i}$ as loss coefficient. For example, a typical value of $r_{i}=2$ for active radio or $r_{i}=4$ for backscatter radio. Now, based on the above model we pose the following lemma.
\begin{lemma}
	\label{lemma:gotoinfinity}
	Any hybrid radio composed of $n$ radios has limited range after which energy consumption per byte is infinite.
	\begin{IEEEproof} $\forall n \lim_{d \to +\infty} B_{\text{rx},i}(d) \to 0 \Rightarrow E_{\text{byte},i} = { E_{\text{tx},i} }/{ B_{\text{rx},i}(d) } \to +\infty$ which completes the proof.
	\end{IEEEproof}
\end{lemma}

\begin{corollary}
	\label{lemma:excludehigh}
	Defining $\mathrm{E}(d) \triangleq \left\{ E_{\text{byte},1}(d),\cdots, E_{\text{byte},n}(d) \right\}$ if $\exists_{E_{\text{byte},i}(d) \in \mathrm{E}(d)} \forall_{E_{\text{byte},j}(d), i \neq j} E_{\text{byte},j}(d) < E_{\text{byte},i}(d), \forall d$, then radio $j$ can be removed from designing a hybrid radio.
\end{corollary}

\begin{corollary}
	\label{lemma:maxrange}
	The maximum range of a system is limited by the radio with the largest range.
\end{corollary}

\begin{corollary}
	\label{lemma:minenergy}
	At distance $d$ the lower bound of the hybrid radio energy consumption per byte is given by the radio with the lowest energy consumption at that distance.
\end{corollary}

\subsubsection{Measurement}
\label{sec:measurements}

To verify this simple analytical model we need to measure the consumed power of each radio as a function of the signal loss. We first introduce the selected hardware for \ac{ble}, \ac{wisp} and finally the measurement setup.

\paragraph{\acl{ble}---Transmitter/Receiver}
\label{sec:motivation/ble}

We selected the Nordic Semiconductor PCA10005 evaluation module with an NRF51822 \ac{ble} \ac{soc}~\cite{nordic2012nrf51822} as \ac{ble} transmitter.
The software used on the \ac{ble} radio is a customized firmware version (source code is available upon request or via~\cite{blisp2015repo}) transmitting only standard advertising messages~\cite{ble2015standard} at a constant rate of 120\,\si{Byte\per\second}=0.96\,\si{\kilo bit\per\second}, which is comparable to 0.65\,\si{\kilo bit\per\second} of~\cite[Sec. III-B]{dementyev2013rfid}.
\ac{ble} has a maximum packet size smaller than the selected payload (i.e. 24\,\si{Byte}) therefore each transmission consists of multiple packets.
A second identical NRF51822 module is used as \ac{ble} receiver---continuously logging advertisement messages send by the \ac{ble} transmitter.

\paragraph{\acl{crfid}---Transmitter/Receiver}
\label{sec:motivation/wisp}

We select \ac{wisp}\,5 as a state-of-the-art \ac{crfid} platform~\cite{wiki2015wisp5}.
The \ac{wisp}\,5 used for experiments has the \ac{rf} energy harvester disabled by desoldering the output pin of the buck converter.
This modification simplifies the energy measurement, as the energy provided to WISP\,5 is not fluctuating in time as in the case of harvested energy.
The \ac{wisp}\,5 firmware is adapted (see again~\cite{blisp2015repo}) to transmit with the same data rate as \ac{ble}.
Again, as in the case of \ac{ble}, since the maximum payload of \ac{wisp}\,5, i.e. 12\,\si{Byte}, is smaller than 120\,\si{Byte} each message consists of multiple packets.
The \ac{rfid} reader is an Impinj Speedway R420~\cite{impinj2014r420}, controlled via SLLURP \ac{llrp} library~\cite{github2015sllurp}, and connected to a panel antenna~\cite{laird2015antenna}.

Based on observations by~\cite[Sec. 4.1]{gummeson2012flit} we have chosen to use the \ac{epcc1g2} \ac{epc} field as our data carrier instead of the \texttt{Read} command.
Using the \ac{epc} field cuts down on the protocol overhead because it halves the amount of roundtrips~\cite[Sec. 6.3.2.12.3]{epcglobal2013gen2}.
According to~\cite[Sec. 6.3.2.1.2.2]{epcglobal2013gen2} the length of the \ac{epc} field may be set between zero and thirty one words.
While it is possible to have \ac{wisp} transmit longer \ac{epc} values to reduce the overhead, this increases the probability of corrupted messages~\cite{lettieri1998info}. 

\paragraph{Measurement Setup}
\label{sec:motivation/method}

We measure energy per byte at the receiver (separately for \ac{ble} and \ac{wisp}) as a function of signal attenuation.
This is realized with two signal attenuators~\cite{jfw2015attenuators} connected in series.
These attenuators limit the signals bi-directionally, and therefore both uplink and downlink are attenuated at the same time.
Both \ac{ble} transmit/receive evaluation boards used are equipped with an antenna connector allowing attenuators to be inserted directly into the transmission channel.
\ac{wisp}, on the other hand, does not provide such an antenna connector and therefore it is positioned at a fixed distance of 50\,\si{\centi\metre} from the interrogator antenna which is then connected to the \ac{rfid} interrogator via the attenuators.

The \ac{ble} module~\cite{nordic2012nrf51822} has an uncalibrated transmission power setting via the \acs{api} of the S110/S120 (transmitter/receiver, respectively) softdevice.
The highest (4\,\si{\decibel m}) and lowest (--30\,\si{\decibel m}) transmission power are tested.
The \ac{rfid} reader is tested at its maximum transmission power (32.5\,\si{\decibel m}).

We measure the power consumption of both radios using a self developed, buffered, differential, sensing circuit monitoring the voltage drop over a 100\,\si{\ohm} shunt resistor in series with the \ac{dut}.
This circuit is coupled to a Tektronix MDO4054B\mbox{--}3 oscilloscope~\cite{tektronix2015mdo4000} to measure power over time which is used to calculate the energy consumption.
Schematics of this device are available upon request or at~\cite{blisp2015repo}.

\subsubsection{Measurement Results}
\label{sec:motivation/results}

\begin{figure}
	\centering
	\includegraphics[width=\columnwidth]{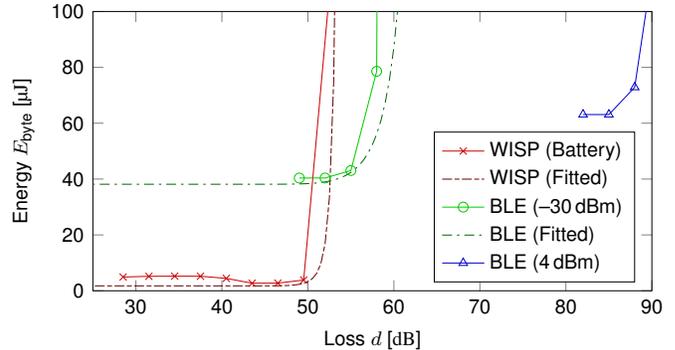}
	\vspace{-25pt}
	\caption{\textbf{Energy per byte over distance for \acs{wisp} and \acs{ble}.}
	The dashed data points are extrapolated, the constant power consumption for the \acs{ble} radio and all data being received, yields constant energy per byte.
	Fitted plots are based upon~(\ref{eq:rx_bytes}).
	Parameters for fitted \acs{wisp} curve: $a_{i} = 30$, $d_{i} = 4$, $E_{\text{tx},i} = (L + H) 5\,\si{\micro\joule}$ with $L = 96$ and $H = 320$.
	Parameters for fitted \acs{ble} curve: $a_{i} = 87$, $d_{i} = 2$, $E_{\text{tx},i} = (L + H) 21\,\si{\micro\joule}$, $L$ and $H$ are equal to \acs{wisp}.
	}
	\label{fig:power_distance}
\end{figure}

The relationship between energy per byte and signal loss, as measured for both active and backscatter radio and complementary fitted plots, is shown in~\Cref{fig:power_distance}.
As expected, the \ac{wisp}---while more energy efficient in good channel conditions---also has a shorter range of operation.
Instead of a gradual increase in energy consumption per received byte, at one point the energy per byte metric starts to rapidly increase for both platforms.
This ``brick wall'' effect~\cite[Sec. V]{lettieri1998info} is caused by an increase in bit errors, causing whole packet loss and therefore requiring more transmit attempts per successfully received byte.

\subsection{Do Alternatives Exist to Hybrid Active/Backscatter Radio?}
\label{sec:motivation/alternatives}

The question remains of whether, in the light of this observation, the hybrid radio platform is the only solution which improves energy efficiency and transmission range of \ac{crfid}. We review the alternatives and provide our answer.

\subsubsection{Low Power Active Radio with Battery}
\label{sec:alternatives/blebatt}

The simplest alternative to the hybrid platform would be a connection of a sufficiently large battery and \ac{ble} radio.

\textbf{Limitation:}
Unfortunately, all batteries will eventually deplete, leading to expensive battery (or even whole device) replacement.
For battery replacement, the device must be physically accessible, as it is impossible to wirelessly restore the energy level of the empty battery without an energy harvester.

\subsubsection{Power Harvester with Active Radio}
\label{sec:alternatives/harvester}

Wireless \ac{rf} power harvesters solve the physical accessibility and battery constraint.

\textbf{Limitation:}
Inefficiencies in \ac{rf} power harvesters, energy storage, energy conversion, and energy transmission through \ac{rf} waves, mean that no power harvester and active radio combination will be as energy efficient as a backscatter radio.

\subsubsection{Backscatter Radio with Improved Channel Coding}
\label{sec:alternatives/coding}

The operational reliability and robustness of communication of \acp{crfid} could be improved by adding a more extensive channel coding mechanism.
For example: \ac{wisp} is currently limited to the FM0 coding~\cite[Sec. 6.3.1.3.2.1]{epcglobal2013gen2}, in which each bit is represented by one signal alternation for each symbol.
Miller coding methods~\cite[Sec. 6.3.1.3.2.3]{epcglobal2013gen2} have redundant alternations within each symbol, reducing the possibility of lost messages.

\textbf{Limitation:}
Channel coding would make \ac{crfid} more robust (i.e. shift the \ac{wisp} curve to the right in~\Cref{fig:power_distance}), still keeping \ac{crfid} susceptible to reflections and destructive interference.
Finally, we conjecture, this would still not make \ac{wisp} as energy efficient as \ac{ble} in a broad attenuation range.

\section{Channel Estimation Methods for Hybrid Active/Backscatter Radio Platforms}
\label{sec:radio-switching}
\label{sec:feedback}

We propose a method of estimating the backscatter channel and use this estimation to select between backscatter and active radio on-the-fly. We start with revising unsuitable solutions.

\subsection{Backscatter Channel Quality Estimation Methods: Review of Unsuitable Solutions}

Because backscatter radios behave differently than active ones, typical channel estimation methods do not directly apply.

\subsubsection{\acs{epcc1g2} Protocol Feedback}
\label{sec:feedback/epc}

The de facto standard method of assessing packet reception rate is to query the receiving party if it indeed received a packet.
Most protocols rely on receive acknowledgments for (all) packets.

\textbf{Limitation:}
Within the \ac{epcc1g2}~\cite{epcglobal2013gen2} protocol there are no standard ways to guarantee the successful reception of {~~}\ac{epc} values transmitted by a tag.
The default method of awaiting an \ac{ack} message for each transmitted data message is therefore not possible.
The exclusion of this functionality is logical for standard \ac{rfid} tags, as they are computationally limited, transmit unchanging identifier, and most likely could not handle retransmissions.
Transmitting data back to a \ac{crfid} also implies that {~~}\ac{crfid} should handle computationally hard, and a protocol-wise large overhead inducing \ac{epcc1g2} \texttt{write} accesses.

\subsubsection{\acs{ble} Protocol Feedback}
\label{sec:feedback/ble}

The more responsive \ac{ble} channel could be used to provide a feedback for the reception of \ac{rfid} packets transmitted by {~~}\ac{crfid}.

\textbf{Limitation:}
The use of a separate radio channel could increase \ac{rfid} reliability because the channels might break down under different circumstances. However, it might also decrease reliability because the \ac{ble} channel might be broken while the \ac{rfid} channel is working.
Practically, including a \ac{ble} radio in receiving mode will also dramatically decrease the energy efficiency of a hybrid platform, as the radio has to listen for an extended (worst case: continuous) time.

\subsubsection{\acs{rssi} Strength Feedback}
\label{sec:feedback/rssi}

Neither {~~}\ac{crfid} hardware nor {~~}\ac{epcc1g2} protocol has a built-in support for \ac{rssi} measurement on the \ac{rfid} transmission.
A coarse method to estimate the vicinity of {~~}\ac{rfid} reader is by measuring the amount of energy harvested by the \ac{crfid}.
If a tag is close to a reader, it is easily possible to harvest energy, while if a tag is far away it would be almost impossible to harvest it.
The \ac{ble} radio has native support for \ac{rssi} measurements on the received messages.

\textbf{Limitation:}
Measuring \ac{rssi} for the signal originating from the interrogator and received by the backscatter radio does not directly correlate with the channel quality for backscatter data (as there is no constructive interference, as explained in~\Cref{sec:introduction/problem}).
While the interrogator knows the \ac{rssi}, the backscatter device cannot reliably determine it.
A \ac{crfid} could query the \ac{rfid} reader for its \ac{rssi} as measured by the reader but this would induce a lot of overhead on both sides. 
I
\ac{ble} should be placed into listening mode in order to retrieve \ac{rssi} values, which is more power consuming than the transmission mode.
Therefore, enabling \ac{ble} only for channel estimation without using it for data transfer is a loss of energy.

\subsection{Proposed Channel Quality Estimation Method}
\label{sec:feedforward}

For the \acs{blisp} system we propose a novel, less standard, way of estimating the channel.
\begin{proposition}
Tracking the number of \ac{epcc1g2} \texttt{RN16} \ac{ack} messages in handshake can be used to estimate the backscatter channel quality.
\begin{IEEEproof}
(Sketch) If {~~}\ac{rfid} interrogator and tag perform a multipart handshake, the backscatter channel is usable to transfer data.
Work of~\cite{zhang2012mobisys} proposed an approach for setting an interrogator to its optimal settings based on both measured \ac{rssi} and packet loss.
However, packet loss-based, estimations can also be performed on the tag instead of the interrogator.
Part of this handshake is the tag sending the reader a random number (\texttt{RN16}), which the reader should acknowledge by an \ac{ack} message containing this random number.
To reach the \ac{ack} both channels (to and from) the \ac{crfid} tag need to be in a state good enough to transmit a payload.
By measuring the number of handshakes and testing this number to be at least the same as the amount of packets we expected to transmit, we are able to estimate quality of the backscatter channel.
\end{IEEEproof}
\end{proposition}

\section{\acs{blisp} Design}
\label{sec:blisp_design}

We are now ready to introduce \ac{blisp}, our hybrid backscatter and active radio platform, to help exploit the main trade-offs as proposed in~\Cref{lemma:gotoinfinity,lemma:excludehigh,lemma:maxrange,lemma:minenergy}. 
The \ac{blisp} infrastructure mainly consists of two parts:
\begin{enumerate*}[label=(\roman*)]
	\item 	a \ac{cots} \acs{rfid} interrogator combined with a \ac{ble} receiver; and
	\item 	our multi-radio sensor node---the \ac{blisp}.
\end{enumerate*}
To provide a flexible platform we opt to combine two readily available radios instead of developing our own, single silicon, platform.

\begin{figure}
	\centering
	\includegraphics[width=0.9\columnwidth]{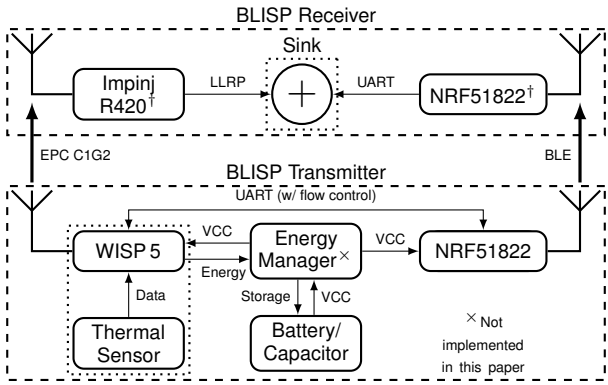}
	\caption{\textbf{Overview of the \acs{blisp} system consisting of one transmitter and one receiver.}
	The temperature sensor providing data is part of {~~}\ac{wisp} but displayed separately for clarity and completeness.
	All displayed connections depict a flow of energy or data and do not directly correspond to physical connections.
	For a detailed description of the physical connections see~\cite{blisp2015repo}.
	\textsuperscript{\dag}For the mobile reader experiments the Impinj R420 is replaced by an MTI MINI ME, and the NRF51822 by the \ac{ble} receiver of a Samsung Galaxy S3.}
	\label{fig:blisp_overview}
\end{figure}

A complete system level overview of the \ac{blisp} is shown in~\Cref{fig:blisp_overview}.
The main design principle behind \ac{blisp} is the absence of any algorithm on the host side: the host only merges the multiple data streams received by the different radios.

\begin{figure}
	\centering
	\subfigure[The top side of the \ac{blisp} with annotations for the most important components.
	Please note that the \acs{wisp} antenna is not fully shown.]{
		\includegraphics[width=0.8\columnwidth]{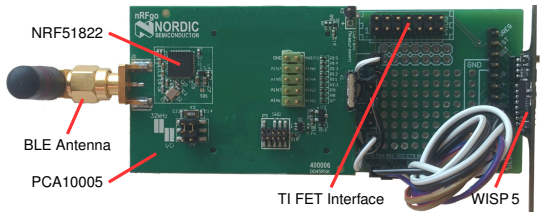}
		\label{fig:blisp_bottom}
	}
	\subfigure[Bottom side of the \ac{blisp}.
	Symbolicaly illustrated directional connections by color, as numbered:
	(1) \textbf{white}: ground;
	(2) \textbf{brown}: clear to send;
	(3) \textbf{red}: power supply;
	(4) \textbf{green}: \ac{wisp} to \ac{ble} serial channel;
	(5) \textbf{orange}: \ac{ble} to \ac{wisp} serial channel (unused);
	(6) \textbf{yellow}: ready to send, and
	(7) \textbf{blue}: power supply.
	]{
		\includegraphics[width=0.8\columnwidth]{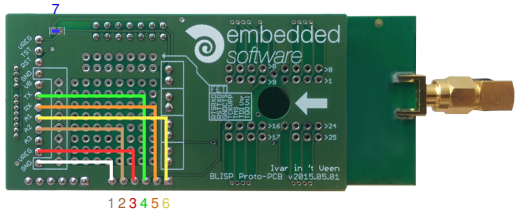}
		\label{fig:blisp_top}
	}
	\caption{\textbf{The \acs{blisp} \ac{pcb} as seen from top (\cref{fig:blisp_top}) and bottom (\cref{fig:blisp_bottom}).}
		\ac{pcb} design files are available upon request or from~\cite{blisp2015repo}.
	}
	\label{fig:blisp_pcb}
\end{figure}

\subsection{\acs{blisp} Hardware Architecture}
\label{sec:blisp_hardware}

The chosen radio modules for this platform are the same as described in~\Cref{sec:measurements}.
{~~}\ac{pcb} has been designed to ease the connection of the two separate radio platforms, see~\Cref{fig:blisp_pcb}.
The \ac{pcb} connects the active and passive radio, and provides means for radio collaboration and energy distribution.

\subsubsection{Active Radio}
\label{sec:blisp/ble}

We use the same NRF51822 \ac{ble} module as described in~\Cref{sec:measurements}.

\subsubsection{Backscatter Radio}
\label{sec:blisp/wisp}

As backscatter radio we also use the same \ac{wisp}\,5 as described in~\Cref{sec:measurements}.

\subsubsection{Radio Collaboration}
\label{sec:blisp/collaboration}

A communication channel is needed to convey desired state information for the active radio and to share sensor values between the two separate radios.
The NRF51822 \ac{ble} module has a silicon bug causing high power consumption by perpetually keeping non-vital microcontroller peripherals enabled~\cite[Id 39]{nordic2013pan2}.
This bug unfortunately affects all conventional (digital) communication channels including \ac{gpio}-interrupts rendering them useless as low power wake-from-sleep devices.
The low power analog comparator peripheral is not affected by this bug, therefore this peripheral is used as wake-up signal enabling the high throughput \ac{uart}. 
The \ac{ble} radio also uses digital output as CTS signal.

\subsubsection{\acs{blisp} Receiver/Sink}
\label{sec:blisp/sink}

The receiving side of \ac{blisp} consists of two receiving radios matching the two transmitting radios on the \ac{blisp}.
In contrast to~\cite{gummeson2010jsac}, the \ac{blisp} receiver is as simple as possible and only merges the data streams from the receiving radios.
Because the host does not make decisions about which radio to use, the \ac{blisp} can switch without synchronization mechanism.
We present two host setups:
\begin{enumerate*}[label=(\roman*)]
	\item 	a fixed receiver; and
	\item 	a mobile smartphone
\end{enumerate*}
setup.

\begin{table}
	\centering
	\scriptsize
	\caption{Setup Parameters of Data Aggregators}
	\vspace{-2mm}
	\begin{tabularx}{\columnwidth}{llXX}
		\toprule
		Component 			& Parameter 		& Mobile \acs{blisp} RX	& Static \acs{blisp} RX \\
		\midrule
		Host Device 		& Model 			& Samsung Galaxy S3		& Lenovo T530 \\
							& Software		 	& Android\,4.3 			& Linux 3.13.0 \\
		\acs{rfid} Reader 	& Model 			& MTI MINI ME 			& Impinj R420 \\
							& TX Power 			& 18\,\si{dBm}			& 32.5\,\si{dBm} \\
							& RX Sensitivity	& --84\,\si{dBm} 		& --82\,\si{dBm} \\
							& Antenna Gain 		& 2\,\si{dBi}			& 9\,\si{dBi} \\
							& Link Frequency	& 640\,\si{\kilo\hertz} & 640\,\si{\kilo\hertz} \\
							& Coding 			& FM0 					& FM0 \\
							& Session 			& 2 					& 2 \\
							& Q-value 			& 5 					& n/a \\
							& Duty Cycle		& 100\% 				& 100\% \\
		\acs{ble} Receiver 	& Model 			& Samsung Galaxy S3		& Nordic NRF51822 \\
							& Duty Cycle		& 100\% 				& 100\% \\
		\bottomrule
	\end{tabularx}
	\label{tab:mobile_data_aggregator}
\end{table}

\paragraph{Fixed Receiver}
\label{sec:blisp/impinj}

The fixed receiver consists of a host computer with an Ethernet connected Impinj Speedway R420~\cite{impinj2014r420} and an \acs{usb}/\acs{uart} connected Nordic Semiconductor NRF51822~\cite{nordic2012nrf51822}.
This setup is again described in~\Cref{sec:measurements}.

\paragraph{Mobile Receiver}
\label{sec:blisp/minime}

To test \ac{blisp} with a mobile reader, comparing to the fixed reader case, we have a prepared the following setup.
Smartphone is selected as platform for mobile host, which consists of \ac{ble} and \ac{rfid} reader.
We developed an Android application (available upon request or via~\cite{blisp2015repo}) for the smartphone to scan the \ac{ble} channel and log all advertising data originating from the \ac{blisp}.
As a smartphone-attachable \ac{rfid} reader we selected the MTI MINI ME~\cite{mti2015minime}. Based on the low level command set \ac{api} provided by MTI, we log all inventory data.

\begin{figure}
	\centering
	\includegraphics[width=0.75\columnwidth]{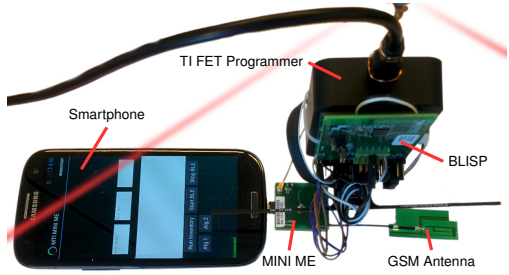}
	\caption{\textbf{Mobile receiver \ac{blisp} test setup.}
	\ac{blisp} and TI FET programmer/power monitor are hanging from an overhead crane (the red diagonal wires)~\cite{catani2015gondola} as described in~\Cref{sec:experiments/setup}.
	MINI ME mobile \ac{rfid} reader is shown without plastic housing, easing connection of a different antenna.
	}
	\label{fig:minime_experiment}
\end{figure}

Unfortunately, the MINI ME can only inventory \ac{wisp} with fixed power supply up to a maximum range of 2\,\si{\centi\metre}.
To increase the inventory range of MINI ME, we replace the embedded antenna with a 2\,\si{dBi} \acsu{gsm} band omnidirectional antenna~\cite{adafruit2015antenna}.
By replacing the antenna, the maximum range is extended to 10\,\si{\centi\metre}.
\Cref{tab:mobile_data_aggregator} shows parameters for the two reader platforms, while
\Cref{fig:minime_experiment} shows the MINI ME reader with \ac{gsm} antenna connected to a smartphone running our application.

\subsection{\acs{blisp} Software Architecture}
\label{sec:blisp_software}

The \ac{wisp} component of the \ac{blisp} software consists of 1700 lines of C code and 1900 lines of assembly code of which 600 lines C and 50 lines assembly were written in the \ac{blisp} development process. The remaining part is based upon~\cite{github2015wisp5}.
The \ac{ble} element consists of a 500 line C coded program and the NRF51822's \ac{api}.
The fixed \ac{blisp} host currently consists of various Bash and Octave scripts with varying lengths.
The mobile host consists of 750 lines of customized Java code.

\subsubsection{\acl{wisp}}
\label{sec:software/wisp}

Because of the low power requirements and therefore our preference for backscatter communication we choose to have {~~}\ac{wisp} acting as master over {~~}\ac{ble} radio.
Between the periodic sensing and transmission rounds {~~}\ac{wisp} is put into a low power state.

For all following experiments \ac{wisp} measures temperature and a timestamp since the startup\footnote{Other possible sensors are the accelerometer, already available on {~~}\ac{wisp}, or any other (low power) electronic sensor.}.
The timestamp is included for evaluation purposes, as this value enables to evaluate the number of missing and/or duplicate packets.
To ensure a constant data stream in case of radio switching the sensor data is periodically shared with the \ac{ble} radio as described in~\Cref{sec:blisp/collaboration}.
The \ac{ble} radio and the \ac{wisp} are both set to have \SI{12}{Byte} payload per message and ten messages are combined into a single transmission.
As the temperature data combined with the timestamp only uses \SI{4}{Byte} the message is padded with \SI{8}{Byte} of constant data.

Because of incompatibilities between {~~}\ac{wisp} and the MINI ME \ac{rfid} reader used for the mobile host experiments the \ac{epcc1g2} \texttt{tag select} mechanism~\cite[Sec. 6.3.2.3]{epcglobal2013gen2} is disabled for all fixed and mobile reader experiments.

\subsubsection{\acl{ble}}
\label{sec:software/ble}

{~~}\ac{ble} module (as decribed in~\Cref{sec:motivation/ble}) is programmed as slave under {~~}\ac{wisp}.
As described in~\Cref{sec:blisp/collaboration} the \ac{ble} radio is periodically awaken by the \ac{wisp} to receive new data.
When not wirelessly transmitting nor receiving (\ac{uart}) data from {~~}\ac{wisp} {~~}\ac{ble} module is put into a low power sleeping state.

\subsubsection{Radio Switching}
\label{sec:software/switching}

The software implements feed-forward channel estimation as proposed in~\Cref{sec:feedforward}.
The circumstances and environmental influences affecting the \ac{rf} performance of the \ac{wisp} might change in a very irregular and most likely unpredictable way.
We therefore propose and evaluate two switching approaches.

\paragraph{Random ($<x$)}
\label{sec:software/random}

Making the switching mechanism depend on past results will decrease the number of unnecessary backscatter channel evaluations, thereby reducing overhead and improving energy efficiency.
Because we assume the environment to have random unpredictable behavior we opt that it does not make sense to include a sophisticated self-learning algorithm.
Our \emph{random backoff} approach implements an ALOHA-inspired random backoff window with a maximum value of $x$.
A low value of $x$ will make the system more responsive while a high value will make the system more stable in the long run.
A pseudo-code representation of this switching algorithm in shown in~\Cref{alg:blisp}.

\begin{algorithm}[tb]
	\caption{\acs{blisp} Control Protocol}
	\label{alg:blisp}
	{
		\fontsize{7}{10}\selectfont
		\begin{algorithmic}[1]
		\Let{$x$}{Maximum backoff window, see~\Cref{sec:software/random}}
		\For{\textsc{Period$_n$}}
			\Let{$a$}{\textsc{\#ACK$_{n-1}$}}
			\Comment{Received ACKs}
			\Let{$r$}{\textsc{\#Frame$_{n-1}$}}
			\Comment{Frames planned to transmit}

			\Let{$\text{WISP}_\text{ok}$}{$\left( a = r \right)$}
			\Comment{Expect ACK for each frame}

			\If{$\text{WISP}_\text{ok}$}
				\Let{backoff}{$0$}
				\Comment{No backoff on success}
			\EndIf

			\If{$0 = \text{backoff}$}
			\Comment{Is (re)try slot?}
				\Let{$\textsc{WISP}_\text{TX}$}{\textsf{true}}
				\Comment{Transmit using WISP}

				\If{$\neg \text{WISP}_\text{ok}$}
					\Let{backoff}{$\Call{$\mathcal{U}$}{0, x}$}
					\Comment{New uniformly random backoff}
				\EndIf

			\Else
				\Let{$\textsc{WISP}_\text{TX}$}{\textsf{false}}
				\Comment{Not transmit using WISP}
				\Let{backoff}{$\text{backoff} - 1$}
				\Comment{Shift backoff}
			\EndIf

			\Let{$\textsc{BLE}_\text{TX}$}{$\neg \text{WISP}_\text{ok}$}
			\Comment{Use BLE if not use WISP}
		\EndFor
		\end{algorithmic}
	}
\end{algorithm}

\paragraph{Na\"ive}
\label{sec:software/naive}

Limiting the maximum random value to zero will generate a constant as-short-as-possible backoff window resulting in the \emph{na\"ive} approach.
This approach (used as a reference) assures that we use \ac{wisp} as much as possible which increases energy efficiency.
At the same time, checking a perpetually broken \ac{wisp} communication channel induces an overhead compared to other maximum backoff window sizes.

\section{Experimental Evaluation}
\label{sec:experiments}

To test the performance of \ac{blisp} we executed the following experiments measuring both goodput and energy consumption.

\subsection{Experiment Setup}
\label{sec:experiments/setup}

Our experimental setup consists of hardware components and methodologies for replicable and traceable measurements.
For this test the \ac{blisp} (built as described in \Cref{sec:blisp_hardware}) was running software as described in \Cref{sec:blisp_software}.

\subsubsection{Hardware}

The measurement and evaluation setup we use for these experiments is based on the setup described in~\Cref{sec:measurements}.
In addition we use an automatic three-dimensional positioning crane~\cite{catani2015gondola} situated in a lab environment to automate the experiments involving a mobile \ac{blisp}.

\subsubsection{Replicability}

According to~\Cref{fig:power_distance} wireless radios have two main ranges of operation:
\begin{enumerate*}[label=(\roman*)]
	\item 	within the first range most of the packets get received and therefore the energy per byte ratio stays rather constant,
	\item 	within the second range almost no packets are received and the energy spend on transmitting a byte therefore increases drastically.
\end{enumerate*}
For the \ac{blisp} performance tests we limit the transmission power and sensitivity of \ac{rfid} reader and define two static positions, one in \ac{wisp}-range and one outside \ac{wisp}-range.
The experiments were performed by placing the \ac{blisp} in the in-range spot, placing the \ac{blisp} in the out-range spot, and alternating the \ac{blisp} location between the in- and out-range positions on a predefined constant time interval (\SI{10}{\second}).
The \ac{ble} radio was in range for all experiments, otherwise the system would fail according to~\Cref{lemma:maxrange}.
The time duration for each experiment was \SI{2}{\minute} and each experiment was repeated five times.
We run baseline experiments with a battery powered \ac{wisp} and a \ac{ble} radio transmitting at \SI{4}{dBm} as used in~\Cref{sec:motivation/efficiency}.

\subsubsection{Data Collection}

In experiments we log the number of received packets for \ac{rfid} and \ac{ble} receivers.
The power consumption is measured by the programmer interface using the EnergyTrace platform\footnote{Because of the limited \ac{api} for the EnergyTrace platform we use synchronously timed screen shots and \ac{ocr} to log the energy measurements for experiments using the EnergyTrace.}~\cite{ti2015energytrace}.
Due to random startup delays of each platform, we match the start and stop of an experiment by asynchronously starting all platforms and logging their state after a fixed (empirically found) delay of \SI{3}{\second}.

\subsection{Static \acs{rfid} Reader Experiment}
\label{sec:experiments/results}

\begin{figure}
	\centering
	\subfigure[Energy per byte comparison for all setups in different scenarios
	]{
		\includegraphics[width=0.85\columnwidth]{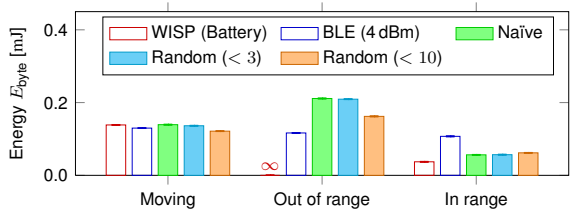}
		\label{fig:perbyte_blisp}
	}
	\subfigure[Number of received messages for all setups in different scenarios
	]{
		\includegraphics[width=0.85\columnwidth]{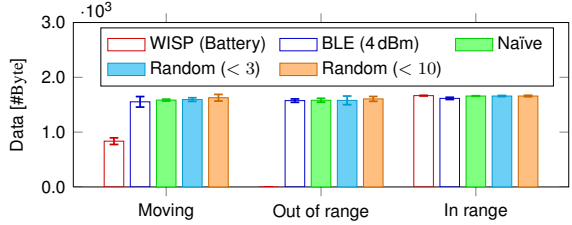}
		\label{fig:rate_blisp}
	}
	\caption{\textbf{Results of the \acs{wisp}, \acs{ble} and \acs{blisp} evaluation using Impinj R420 \acs{rfid} reader.}
	Because of \acs{wisp} not being able to transmit data in the long range, see~\Cref{fig:rate_blisp}, effectively wasting energy, the energy per byte is infinite for this situation, see~\Cref{fig:perbyte_blisp}.
	We show the only \acs{wisp}, only \acs{ble}, na\"ive \acs{blisp} and random \acs{blisp} for random backoff windows up to three and ten slots.
	These experiments have been normalized to \emph{unique} messages eliminating messages transmitted by both radios around switching moments.
	}
	\label{fig:blisp_results}
\end{figure}

Measurements of energy per byte and transferred data, are shown in~\Cref{fig:perbyte_blisp} and~\Cref{fig:rate_blisp}, respectively. Due to normalization to unique messages, the values in~\Cref{fig:perbyte_blisp} are around ten times larger than the ones shown in~\Cref{fig:power_distance}.

Our experiments show that \ac{blisp} increases goodput almost infinitely in the long range compared to \ac{wisp} (see~\Cref{lemma:gotoinfinity}) while not severely increasing power consumption over {~~}\ac{wisp} in the short range.
On the other hand \ac{blisp} almost halves energy consumption in the short range compared to a normal \ac{ble} radio while for Random ($x<10$) increasing energy consumption by $\approx$25\% on the long range. For the remaining two switching methods this difference is much larger. This is presumably caused by the amount of unneeded channel sensing operations and the overhead of redundant micro-controllers.
As we add a mobility to the experiment we see \ac{wisp} loosing a share of messages corresponding to the relative out of range time, this increases the energy per byte to the same level as the active \ac{ble} radio which is able to transfer data in all positions.
The combined system cannot be more energy efficient than the most efficient radio for a certain position (see~\Cref{lemma:minenergy}).

For an uniformly distributes in-/out-range mobility pattern the energy profit the \ac{blisp} has over the \ac{ble} radio in short range and the energy cost in the long range zero out.
\ac{blisp} improves energy efficiency and throughput for situations in which the \ac{wisp} can be used for half of the time.

\subsection{Mobile \acs{rfid} Reader Experiment}
\label{sec:experiments/minime}

\begin{figure}
	\centering
	\subfigure[Energy per byte comparison for all setups in different scenarios
	]{
		\includegraphics[width=0.85\columnwidth]{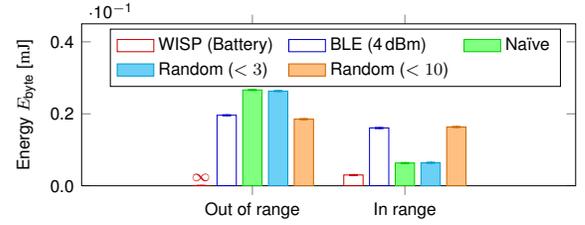}
		\label{fig:perbyte_minime}
	}
	%
	%
	\subfigure[Received messages per radio (\acs{ble} or \acs{wisp}).
		Note: dark (top) part of bars---\acs{ble} messages, light (bottom) part---\acs{wisp} messages
	]{
		\includegraphics[width=0.85\columnwidth]{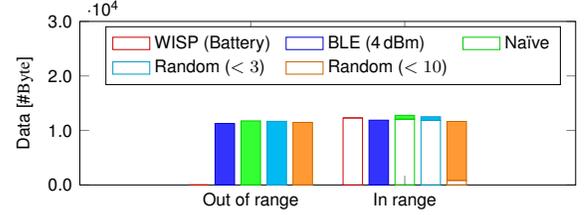}
		\label{fig:rawrate_minime}
	}
	\caption{\textbf{Results of the \acs{wisp}, \acs{ble} and \acs{blisp} evaluation using MiniMe \acs{rfid} reader.}
	Again, in the long range the MINI ME reader is not able to receive data transmitted by the \acs{wisp}. 
	\Cref{fig:rawrate_minime} shows the distribution of received messages per radio for completeness of the illustration.
	Because of the limited logging capabilities on the smartphone the number of messages is not normalized to number of unique messages.
	}
	\label{fig:minime_results}
\end{figure}

The experiment setup parameters on the \ac{blisp} side is the same as using fixed \ac{rfid} reader in~\Cref{sec:experiments/setup}.
The detail setup parameters of mobile data aggregator are as in~\Cref{tab:mobile_data_aggregator}.

Results for the mobile host experiments as shown in~\Cref{fig:minime_results} show comparable results among \ac{wisp}, \ac{ble} and \ac{blisp} compared with fixed reader experiments from~\Cref{sec:experiments/results}.
The relative improvement from \ac{ble} to \ac{wisp} and \emph{na\"ive}-\ac{blisp} using a mobile reader is even larger while in-range.
This relative improvement is mainly because the performance of the smartphone's \ac{ble} module has worse performance than the NRF51822 receiver.
Interestingly, for in-range measurements, a large backoff window shows worse performance than the \emph{na\"ive} and small backoff experiments.
We suspect that this is caused by the hardware limitation of MINI ME.
Based on our experiments, the MINI ME reader has trouble with rapidly moving, or only shortly available, \ac{rfid} tags.
Fortunately, the \ac{blisp} algorithm detects the failing \ac{rfid} reader and correctly enables the \ac{ble} radio which results in continuous data availability.

\section{Limitations and Future Work}
\label{sec:future}

We list the limitations and action items for future work related to hybrid active/passive radio platforms:

\begin{enumerate}
	\item 	\textbf{Improving platform switching mechanism:} Non-predictable mobility patterns require further research on learning mechanism to select the best backoff parameter $x$ of~\Cref{alg:blisp}, or the complete redesign thereof.
	\item 	\textbf{Reducing micro-controller overhead:} The current \ac{blisp} is built using two separate radio modules and therefore two micro-controllers.
One micro-controller is a better approach, reducing energy consumption of \ac{blisp}.
	\item 	\textbf{Extending to beyond two radio platforms:} 
By~\Cref{lemma:maxrange} and~\Cref{lemma:minenergy} adding radios with heterogenous characteristics to a hybrid system will increase the performance of \ac{blisp}, requiring research on radio selection.
\end{enumerate}

\acresetwithnoexpand

\section{Conclusion}
\label{sec:conclusions}

In this paper we design, implement, and evaluate a hybrid radio platform composed of \acf{wisp} and \acf{ble}, denoted as \acs{blisp}.
Through experiments we show that \acs{blisp}, in situations in which this hybrid platform stays within the reception region of the lowest power radio, i.e. \ac{wisp}, the energy efficiency is improved compared to \ac{ble}. At the same time the reliability of \acs{blisp} is larger than the reliability of \ac{wisp} alone when \acs{blisp} moves frequently away from the \ac{rfid} reader range.


\end{document}